\documentclass[a4paper,12pt]{article}
\usepackage[utf8]{inputenc}
\usepackage{amssymb,amsmath}
\usepackage{graphicx}
\usepackage{hyperref}
\title{Geometrothermodynamics of BTZ black hole in new massive gravity}
\author{Jishnu Suresh$^{1}$ and V. C. Kuriakose$^2$ \\
\vspace{0.2in} \\
 Department of Physics, \\
Cochin University of Science and Technology, 
\\{Kochi-22, India}.\\
\vspace{0.1in} \\
E-mail: $^1$ jishnusuresh@cusat.ac.in \\
$^2$vck@cusat.ac.in\\}

\begin{document}
\date{}
\maketitle

\begin{abstract}
We investigate the thermodynamics as well as thermodynamic geometry of chargeless BTZ black hole solution in new massive gravity. Phase structure and thermodynamic stability of the system is analyzed using the Geometrothermodynamic approach. The phase transition between BTZ black hole space time and thermal AdS$_{3}$  soliton is studied using the same approach and the existence of a second order phase transition is examined. 
\end{abstract}

\section{Introduction}
General relativity survived all most all kinds of theoretical and observational challenges.
It is an accurate theoretical model for macroscopic gravitational interactions but still now a quantum formulation of GR is not successful. At the solar scale, the theoretical predictions of GR has passed all the experimental tests \cite{will}. But in the very 
large distance range, more complications occurred due to the disagreement of theoretical predictions with that of observed 
nature of gravity at cosmological scales. These contradictions include the cosmological constant problem \cite{weinberg}, the presence of 
dark matter component in the universe \cite{zwicky} and the late time accelerated expansion of the universe \cite{perlmutter}. 
Hence it is reasonable to think and investigate the possibility of modifying Einstein's GR in order to incorporate these observations. 
We know that GR describes nonlinear self interactions of a massless spin 2 excitations. So the most natural way of modifications would be 
by adding a mass term for the spin 2 field. As a result the modified theory would explain the nonlinear interactions of a massive spin 2 field, and 
this theory is known as Massive gravity. In 1939, Fierz and Pauli \cite{fp} considered modification by adding mass to a linearized theory of gravity.
The proposed theory was unique in such a way that there were only a single way to add a mass term so that the theory becomes physically significant. 
Later in two independent articles, van Dam and Veltman and Zakharov claimed that Fierz-Pauli theory could not converge to Einstein's theory
in the zero mass limit \cite{vd,zak}.  Later in 1972, Vainshtein introduced a new mechanism to overturn the above mentioned 
vDVZ discontinuity \cite{vain}. In this mechanism he considered the full non linear formulation of massive gravity and as a result in the 
zero mass limit Einstein's original equations are retrieved. In the same year, Boulware and Deser proved that any non linear massive gravity theory
which uses the Vainshtein's mechanism would contain a 'ghost' \cite{bd1,bd2}, known as the BD ghost. The presence of BD ghost remained as an unsolved 
problem until 2010, when de Rham, Gabadadze and Tolley (dRGT) proposed the first non linear completion of the FP theory 
free of BD ghost instability \cite{drgt1,drgt2}. They showed that the potential to be ghost free up to the quartic
order in perturbation and to all orders in decoupling limit, and as a result of this many extensions of this theory 
are discovered \cite{has1,has2,has3,rosen1,rosen2,aragone,isham,salam}. Besides these extensions, alternative theories with massive graviton have also 
been under rigorous investigation. These theories include the DGP model \cite{dgp1,dgp2}, Kaluza - Klein scenarios \cite{kk,kkk}, 
New massive gravity \cite{bht} and Topological Massive gravity \cite{deser1,deser2}.

Black holes have been a mystery and a subject of interest since they have been identified as thermodynamic objects from the 
pioneering works of Hawking and Bekenstein \cite{hawk,bek1,bek2}. By mapping the gravity system in to a thermodynamic system, the four 
laws of black hole mechanics were introduced in \cite{bardeen}. 
The study on thermodynamics
of black holes will eventually shed lights in to the better understanding of the gravity theory. Hence extensive efforts have been devoted for the study of thermodynamic properties of black holes.
The black hole thermodynamic studies, that includes phase transitions \cite{hp}, which are entropically driven in Einstein's theory, 
but depends on other parameters in modified 
theories, are of prime interest in this paper. 
One can study the thermodynamic properties of the black hole by incorporating differential geometry ideas in to black hole thermodynamics.
Geometrothermodynamics \cite{quevedo} is the newest candidate in this direction, where the invariance of classical thermodynamics 
under a change of thermodynamic potential is taken in to account. For that, Legendre invariant metric of the phase space is constructed and it will
exactly reproduce the thermodynamic interactions of the present system. So one can get an exact explanation of the phase transition
picture of the black hole system and hence more insight of the gravitational theories.

In the recently proposed three dimensional higher derivative gravity model, the so called new massive gravity (NMG),
higher curvature terms (fourth order)
are added to the usual Einstein-Hilbert action. Unlike the topological massive gravity, the parity is preserved in 
this NMG and it has two propagating bulk degrees of freedom corresponding to the massive graviton modes with spin $\pm 2$. 
The BTZ black hole solutions in this massive
gravity was proposed in \cite{clement}. Many studies have been done regarding the thermodynamic as well as the phase transition aspects of the BTZ black hole solution \cite{btzaa,btza,btzb,btzc,btzd,btze}. The purpose of this paper is to explore the thermodynamics of the BTZ black hole solution in new massive gravity through the geometric point of view using  Geometrothermodynamic method and to understand the phase transition picture as well as the thermodynamic interactions of the same system.

A brief outline of the paper is as follows: in Section 2, We explain the BTZ black hole solutions in new massive gravity. We
calculate the relevant thermodynamic quantities like the horizon
temperature, mass, entropy and the specific heat and the Geometrothermodynamic method for analyzing 
the phase transition picture are depicted in Section 3. The results are summarized in Section 4.

\section{BTZ black hole in New massive gravity and Thermodynamics}
Three dimensional higher derivative gravity model, the New Massive Gravity (NMG) was proposed by Bergshoeff, Hohm and Townsend in 2009 \cite{bht}. 
Their action can be written as a higher curvature term added to the usual Einstein-Hilbert action,
\begin{eqnarray}
 S_{\rm NMG} &=&S_{\rm EH}+S_{\rm R },  \\
 \label{NEH} S_{\rm EH}&=& \frac{1}{16\pi G} \int d^3x \sqrt{-g}~
  (R-2\lambda) \\
\label{NFO} S_{\rm R }&=&-\frac{1}{16\pi Gm^2} \int d^3x
            \sqrt{-g}~\Big(R_{\mu\nu}R^{\mu\nu}-\frac{3}{8}R^2\Big),
            \label{NMGAct}
\end{eqnarray}
where $m^2$ is a mass parameter with mass dimension and $G$ is a three dimensional Newton constant. The equation of motion is given by, 
\begin{equation}
 G_{\mu\nu}+\lambda
g_{\mu\nu}-\frac{1}{2m^2}K_{\mu\nu}=0
\end{equation}
where $G_{\mu\nu}$ is the Einstein tensor given by
$$G_{\mu\nu}=R_{\mu\nu}-\frac{1}{2}g_{\mu\nu}R$$ and 
\begin{eqnarray}
  K_{\mu\nu}&=&2\square R_{\mu\nu}-\frac{1}{2}\nabla_\mu \nabla_\nu R-\frac{\square{}R}{2}g_{\mu\nu}
        +4R_{\mu\rho\nu\sigma}R^{\rho\sigma}\nonumber \\
        &-&\frac{3R}{2}R_{\mu\nu}-R^2_{\rho\sigma}g_{\mu\nu}
         +\frac{3R^2}{8}g_{\mu\nu}.
\end{eqnarray}
Now let us choose the parameters in such a way that,  we will end up with BTZ black hole solution \cite{clement}. For that, we write
\begin{equation}
 m^2=\frac{\Lambda^2}{4(-\lambda+\Lambda)}, \hspace{2cm} \Lambda=- \frac{1}{l^2},
\end{equation}
where $\Lambda$ is the cosmological constant.
From this, one can write the BTZ solution as ,
\begin{equation}
 ds^2_{\rm BTZ}=-f(r)dt^2+\frac{dr^2}{f(r)}+r^2d\phi^2,
\end{equation}
\begin{equation}
 f(r)=-M+\frac{r^2}{\ell^2}
 \label{metric_btz}
\end{equation}
where $M$ is the integration constant corresponding to the ADM mass. 
Horizon radius $r_{+}$ can be determined using the condition, $f(r_{+})=0$. Then the horizon is located at,
\begin{equation}
 r_{+}=l \sqrt{M}.
\end{equation}
Now one can calculate the thermodynamic quantities using the above relations. ADM mass of the black hole can be written as,
\begin{equation}
 M=\frac{r_{+} ^2}{l^2}.
\end{equation}
Hawking temperature is obtained from the relation, $T=\frac{\kappa}{2 \pi}$, as,
\begin{equation}
 T_H= \frac{\sqrt{M}}{2 \pi l}.
\end{equation}
From the ADM mass, entropy of the BTZ black hole can be calculated using either Cardy formula or Wald's formula. We adopt Wald's method to calculate
the entropy as,
\begin{equation}
  S= 2\pi \oint_h  dx \sqrt{\gamma} \frac{\delta \mathcal{L}}{\delta \mathcal{R} _{\mu \nu \rho \sigma}} \epsilon_{\mu \nu} \epsilon_{\rho \sigma},
\end{equation}
where $h$ is the spatial cross section of the event horizon, $\gamma$ is the determinant of the induced metric on $h$, $\mathcal{L}$ is the Lagrangian
in the action (\ref{NMGAct}) and $\epsilon_{\mu \nu}$ is the binormal to $h$. 
So we obtain it as,
\begin{equation}
 S=\frac{\pi  r_{+}}{2G} \Big( 1- \frac{1}{2 m^2 l^2} \Big).
 \label{wald_entropy}
\end{equation}
 It is interesting to note that the entropy (\ref{wald_entropy}) is simply Bekenstein-Hawking entropy renormalized by a factor.
From the above relation one can conclude that in order to get a non-zero central charge we need to rely on the condition $m^2 l^2 \geq 1/2$. 
We will explore this condition in details in the preceding sections. 
The thermodynamic quantities of the BTZ black hole, heat capacity and on shell free energy can be calculated using Abott-Deser-Tekin (ADT) approach \cite{adt1,adt2,adt3}. 
Using this method one can obtain, on shell free energy as,
\begin{equation}
 F^{\mbox{on}} _{\mbox{bh}} = \frac{-M}{8G} \Big( 1- \frac{1}{2 m^2 l^2} \Big),
\end{equation}
thermodynamic energy as,
\begin{equation}
 E = \frac{M}{8G} \Big( 1- \frac{1}{2 m^2 l^2} \Big),
\end{equation}
and the heat capacity as,
\begin{equation}
 C=\frac{\partial E}{\partial T}=\frac{\pi l \sqrt{M}}{2G} \Big( 1- \frac{1}{2 m^2 l^2} \Big).
 \label{heatcapacity}
\end{equation}
\begin{figure}
\includegraphics{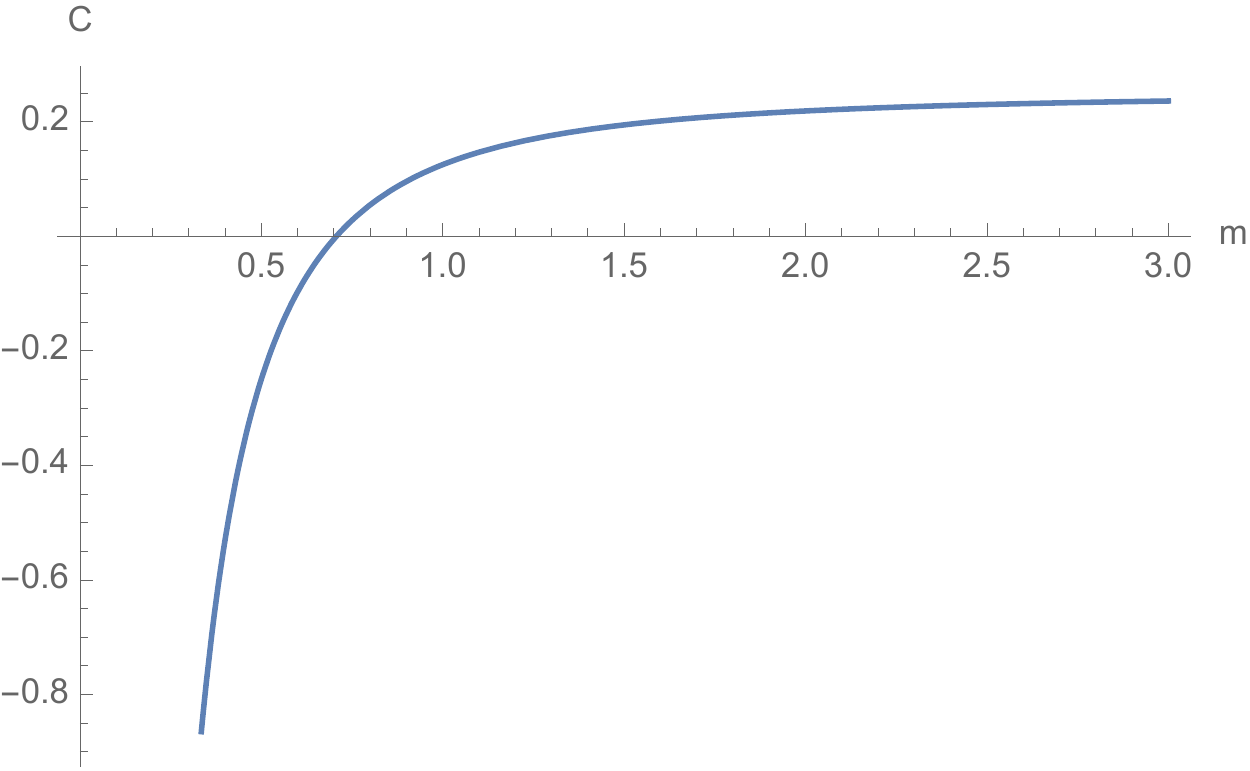}
\caption{Variation of heat capacity of the BTZ black hole against the changes in mass of the graviton with $l=1$, $G=1$ and $M=1$.}
\label{spec_btz}
\end{figure}
In order to investigate the thermodynamic stability of the black hole space time, we will further explore the above equation. 
Therefore, we have plotted the variation of heat capacity with entropy of the BTZ black hole in fig (\ref{spec_btz}).  From this figure 
it is evident that, there exist a point where the heat capacity changes sign continuously, showing a transition between thermodynamically
stable and unstable phases. We can see that there exists one such point, where heat capacity changes from thermodynamically unstable phase to stable phase in a continuous manner rather than in a usual discontinuous way as seen in many cases where black hole exhibits second order phase transition. From (\ref{heatcapacity}), the BTZ black hole is stable when $m^2 l^2  \geq \frac{1}{2}$ and unstable when $0<m^2 l^2<\frac{1}{2}$.

\begin{figure}
\includegraphics{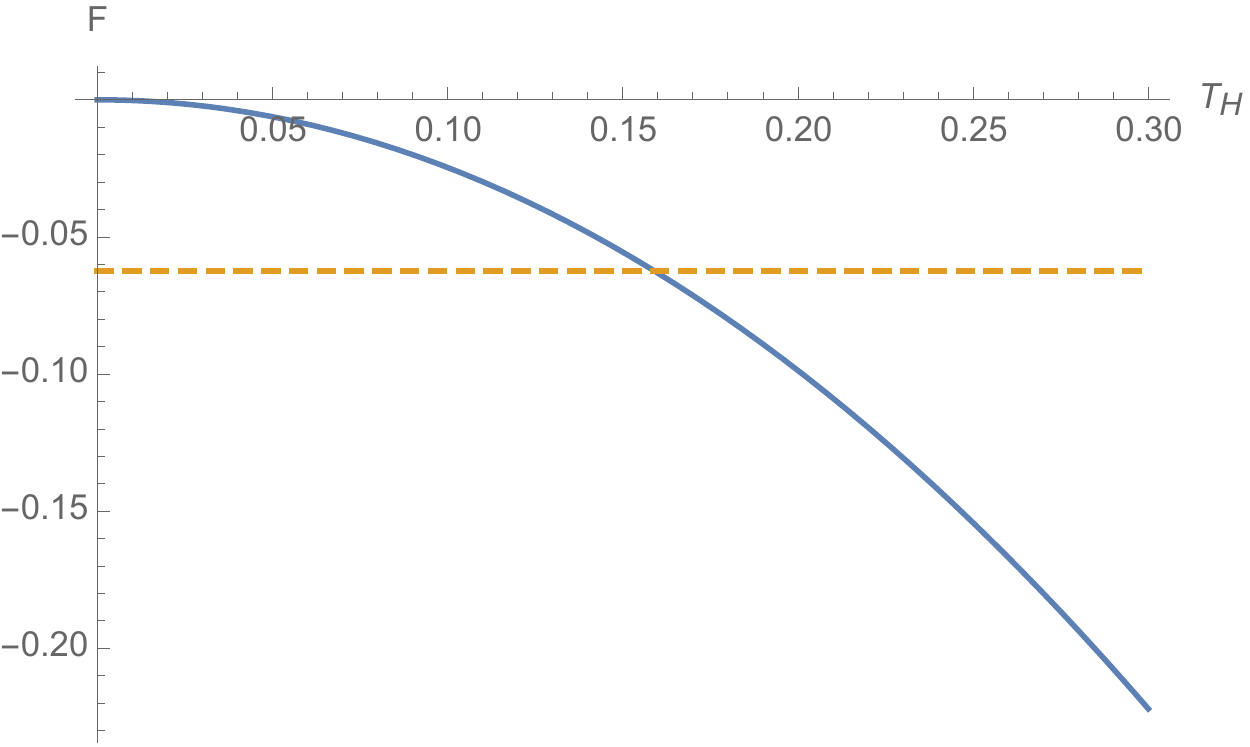}
\caption{variation of free energy of BTZ black hole and thermal solitons in AdS$_{3}$ against the changes in temperature. The solid line represent the behaviour of BTZ black hole, while the dashed line represents the thermal soliton case. }
\label{free_combi}
\end{figure}

We know that, the Banados-Teitelboim-Zanelli (BTZ) black hole system \cite{btz} , there are two distinct solutions of the BTZ black hole of 
$M\geq0$ and the thermal soliton of the global AdS$_{3}$ whose mass is $M = -1 $ \cite{solit1,solit2,solit3}.  Since we have already considered the black hole case, we will dig in to the thermodynamics of thermal soliton of the global AdS$_{3}$  with mass, $M=-1$. In this case, the free energy of the thermal solitons can be calculated as,
\begin{equation}
F^{\mbox{on}} _{\mbox{sol}} = \frac{-1}{8G} \Big( 1- \frac{1}{2 m^2 l^2} \Big) ,
\end{equation}
 In order to analyze the phase transition between BTZ black hole and thermal soliton, let us plot the free energies of both black hole and soliton as a function of temperature. The variation of the same is plotted in fig (\ref{free_combi}). 
From the figures it is evident that the BTZ black hole undergoes a phase transition to thermal soliton of the global AdS$_{3}$ at the critical temperature given by,
\begin{equation}
T_c = \frac{1}{2\pi l}
\end{equation}
From fig(\ref{free_combi}), a phase transition may occur at $T=T_c$ between BTZ black hole and thermal AdS$_{3}$  soliton. From the same figure we can see that, for $T<T_c$, free energy of thermal soliton is lower than that of the black hole. So it can be inferred that the thermal soliton is more probable below the critical temperature. On the other hand for $T>T_c$,  the BTZ black hole is more probable than the thermal soliton. To analyze the phase transition behaviour further, we will consider the  idea of Geometrothermodynamics in the next section.

\section{Review of Geometrothermodynamics and Curvature singularity analysis of BTZ black hole}
Geometrothermodynamics is the new method of describing the phase transitions of a thermodynamic system by incorporating the ideas of differential geometry and Legendre invariance. In this method one can consider the curvature singularities as the phase transition points and hence system interactions can be explained well.  The main constituent of GTD is a $(2n+1)$ dimensional manifold, usually referred to as thermodynamic phase space $(\mathcal{T})$. This phase space is constructed using a set of coordinates $Z^a = \{ \Phi, E^a, I^a \}$, where $E^a$ and $I^a$ are extensive and their corresponding dual intensive variables. Here, $A=0,....., 2n$ and $a=1,...,n$, then the Gibbs 1-form will be,
\begin{equation}
 \Theta = d\Phi- \delta_{ab} I^a dE^b, ~~ \delta_{ab}=\mbox{diag}(1,1,,..,1),
\end{equation}
with summation over the repeated indices. Now if $\mathcal{T}$ is differentiable and $\Theta$ satisfies the condition
$\Theta \wedge (\mbox{d} \Theta)^{n} \neq 0$, then $(\mathcal{T},\Theta)$ can be called as a contact manifold. 
Now consider an $n$ dimensional manifold $\mathcal{E}$, which is a sub manifold of  $\mathcal{T}$, i.e., $\mathcal{E} \subset \mathcal{T}$, 
which can be defined using the extensive thermodynamic variables $E^a$.
This equilibrium manifold $\mathcal{E}$ can be realized by considering a smooth harmonic mapping 
 $\varphi : \mathcal{E} \rightarrow \mathcal{T}$,
  \begin{equation}
  \varphi : \mathcal{E^a} \rightarrow \{ Z^a (E^a)\}=\{ \Phi(E^a),E^a , I^a(E^a)  \},
 \end{equation}
and the condition $\mathcal{E} \subset \mathcal{T}$, can be realized using the condition,
\begin{equation}
\varphi^* (\Theta) = \varphi^* (\mbox{d}\Phi- \delta_{ab} I^a dE^b) =0,
\label{pullback}
\end{equation}
where $\varphi^*$ is the pullback. From the relation (\ref{pullback}) one can easily deduce the 
condition for thermodynamic equilibrium as,
\begin{equation}
 \frac{\partial \Phi}{\partial E^a} = \delta_{ab} I^b.
\end{equation}
Considering the equilibrium manifold $\mathcal{E}$ and using (\ref{pullback}), the first law of thermodynamics can
be written as,
\begin{equation}
\varphi^*(\Theta)=\varphi^*(d\Phi -\delta_{ab}\, I^a\, dE^b) = 0\ .
\end{equation}
The harmonic map $\varphi$ demands the existence of the function $\Phi=\Phi(E^a)$, which is commonly known as the fundamental 
equation in classical thermodynamics from which one can deduce all the equation of states corresponding to that system. From this
fundamental equation, one can write the second law of thermodynamics as,
\begin{equation}
 \pm \frac{\partial^2 \Phi}{\partial E^a  \partial E^b} \geq 0 ,
\end{equation}
also known as the convexity condition. In the above equation, the sign depends on the choice of 
the thermodynamic potential. For example, if one chooses 
$\Phi$ as entropy, then the sign becomes positive and it becomes negative when the potential is chosen 
to be the internal energy. 
Now let us consider a Riemannian metric
$G$ on $\mathcal{T}$, which must be invariant with respect to Legendre transformations. Then the Riemannian 
contact manifold can be defined
as the set $(\mathcal{T},\Theta,G)$ and the equilibrium manifold can be written as a sub
manifold of $\mathcal{T}$, i.e., $\mathcal{E} \subset \mathcal{T}$.
This sub manifold satisfies the above discussed pull back condition \cite{jackle}. 
The non-degenerate metric $G$ and the thermodynamic metric $g$ can be written as,
\begin{equation}
 G=(d\Phi - \delta_{ab} I^a d E^b)^2 +(\delta_{ab} E^a I^b)(\eta_{cd} d E^c d I^d),
\end{equation}
and,
\begin{equation}
  g^Q=\varphi^*(G)=\left(E^{c}\frac{\partial{\Phi}}{\partial{E^{c}}}\right)
\left(\eta_{ab}\delta^{bc}\frac{\partial^{2}\Phi}{\partial {E^{c}}\partial{E^{d}}} dE^a dE^d \right),
\label{quevedo metric}
\end{equation}
with $\eta_{ab}$=diag(-1,1,1,..,1) and this metric is Legendre invariant because of the invariance of the Gibbs 1-form.
So by calculating the curvature scalar of the GTD metric(\ref{quevedo metric}), one can use GTD as a method to investigate
the phase transition structure of the black hole system.

Now we will apply this formalism in to BTZ black hole in new massive gravity.  For that, let us consider a $5$ dimensional
thermodynamic phase space, constituted by the coordinates $Z^a = \{ M, S, l, T, \alpha \}$, where $S,l$ are extensive variables
while $T$ and $\alpha$ are their corresponding dual intensive variables.  From the ideas of equilibrium manifold discussed above, one can obtain the GTD metric as,
\[
         g=(SM_{S}+l M_{l})
            \left[ {\begin{array}{cc}
             -M_{SS} & 0  \\
             0 & M_{ ll} \\
             \end{array} } \right].
        \]
Now one can calculate the Legendre invariant scalar curvature corresponding to the above hessian metric in mass representation as,
\begin{equation}
R_{\mbox{GTD}}=\frac{\pi ^4 l^4 \left(m^2 l^2-2\right)^4 \left(8 m^2 l^2-15\right)}{32 G^4 s^4 \left(5
   m^4 l^4-9 m^2 l^2+6\right)^2}
\end{equation}
\begin{figure}
\includegraphics{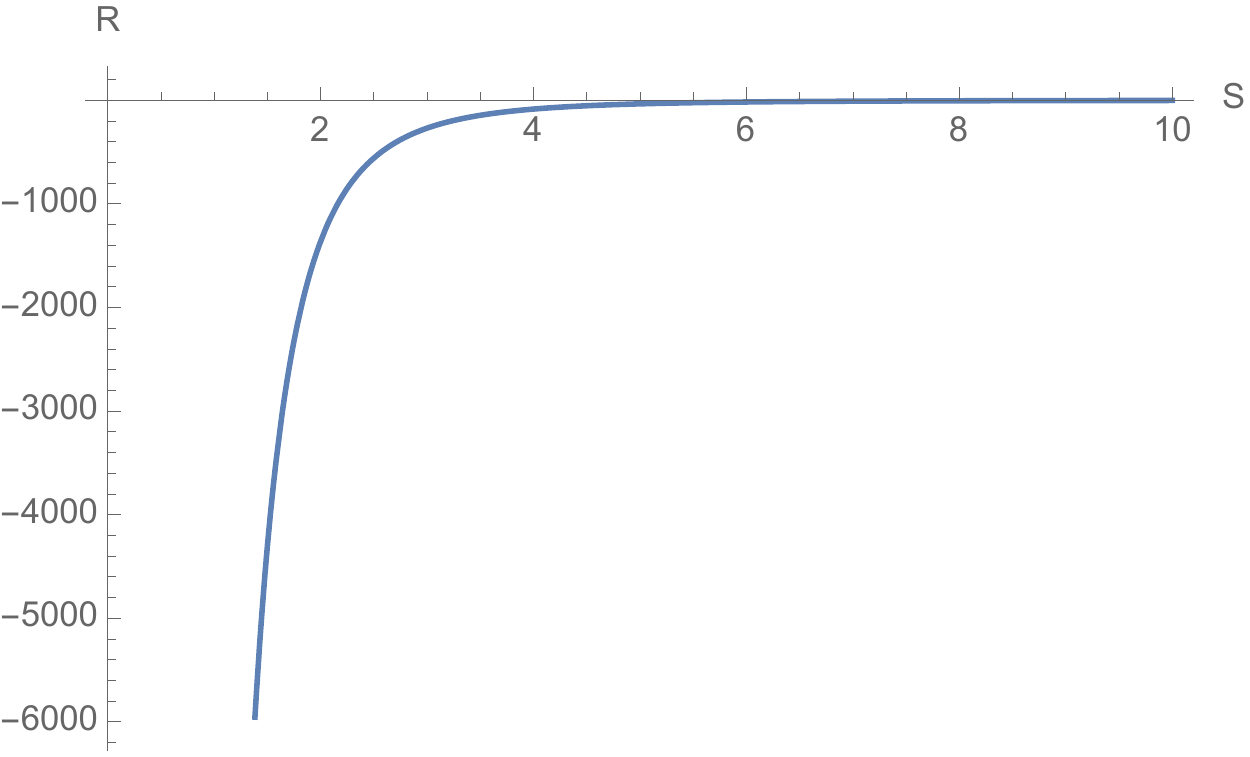}
\caption{Variation of scalar curvature of the BTZ black hole against the changes in entropy with $l=1$, $G=1$ and $M=1$.}
\label{curv_btz}
\end{figure}
We will now explore the thermodynamic behaviour of the system as well as their interactions using this scalar curvature. By plotting the scalar curvature as a function of entropy, the variation is depicted in fig(\ref{curv_btz}).  From this figure it can be inferred that, the particular equilibrium manifold under consideration is a space of negative curvature for any values of entropy or for any values of horizon radius. Hence the scalar curvature corresponding to the BTZ black hole does not possess any discontinuities or zeros. Then we can say that the BTZ space time is free of any thermodynamic curvature singularities. As we have already discussed about heat capacity variation and the non existence of any discontinuities in their variations implies the non existence of second order phase transition. According to GTD formalism, the regular variation of curvature scalar indicates that no (second-order) phase transition occurs. This result does not imply that there is no thermodynamic interactions exists in the case of BTZ black hole, but no second order phase transitions can occur. 

\section{Result and Discussion}
In this work we used the formalism of GTD to construct a thermodynamic equilibrium phase space to study the thermodynamic behaviour of the BTZ black hole solution in new massive gravity. This method shows that the thermodynamic curvature corresponding to BTZ black hole is non zero and free of singularities, indicating the absence of second order phase transition.  
Even though the black hole system shows a continuous transition from a thermodynamically stable to an unstable phase, it is not second order in nature. 

\section*{Acknowledgments}
VCK wishes to acknowledge the Associateship of IUCAA, Pune, India.

\end{document}